\documentclass[%
10pt,%
aps,%
prb,
letterpaper,
twocolumn,%preprint,%
superscriptaddress,%
preprintnumbers,%
floatfix%
]{revtex4}

\usepackage[usenames]{color}
\usepackage{times}
\usepackage{dcolumn}% Align table columns on decimal point
\usepackage{bm}% bold math
\usepackage{ifthen}
\usepackage{amssymb}
\usepackage{amsfonts}
\usepackage{amsmath}
\usepackage[dvips]{graphicx}

\def\figurename{\textbf{\textsf{Figure}}}
\renewcommand \thefigure {\textbf{\textsf{\arabic{figure}}}}

\newcommand\myCaption[2]{\begin{flushleft}\fontsize{7}{10}\selectfont\refstepcounter{figure}%
\noindent\textsf{\figurename}~\fontsize{7}{10}\selectfont\thefigure$\vert$~\fontsize{7}{10}\selectfont{\textsf{\textbf{#1}}}
{\textbf {#2}}\end{flushleft}\small\selectfont}

\newcommand{\pippo}[1]{\centerline{#1}}
\newcommand{\beann} {\begin{eqnarray*}}
\newcommand{\eeann} {\end{eqnarray*}}
\newcommand{\bea} {\begin{eqnarray}}
\newcommand{\eea} {\end{eqnarray}}

\newcommand{\lsb} {\left[}
\newcommand{\rsb} {\right]}

\newcommand{\lcb} {\left\{}
\newcommand{\rcb} {\right\}}

\newcommand{\ImP} {{\rm Im}~}

\newcommand{\g}{^\circ}

\newcommand{\lnothing} {\left.}
\newcommand{\rnothing} {\right.}

\begin{document}
\date{\today}

\title{\huge\bfseries\noindent\sloppy \textsf{Tuning the conductance of a molecular switch}}

\author{\textsf{Miriam del Valle}}
\affiliation{Institute for Theoretical Physics, University of
Regensburg, D-93040 Regensburg, Germany} \affiliation{Departamento
de F\'{\i}sica Te\'{o}rica de la Materia Condensada, Universidad
Aut\'{o}noma de Madrid, E-28049 Madrid, Spain}
\author{\textsf{Rafael Guti{\'e}rrez}}
\affiliation{Institute for Theoretical Physics, University of
Regensburg, D-93040 Regensburg, Germany}
\author{\textsf{Carlos Tejedor}}
\affiliation{Departamento de F\'{\i}sica Te\'{o}rica de la Materia
Condensada, Universidad Aut\'{o}noma de Madrid, E-28049 Madrid,
Spain}
\author{\textsf{Gianaurelio Cuniberti\footnote{Corresponding author}}}
\affiliation{Institute for Theoretical Physics, University of
Regensburg, D-93040 Regensburg, Germany}

\maketitle \small \noindent \textbf{ The ability to control the
conductance of single molecules will have a major impact in
nanoscale
electronics\cite{joachim05,lindsay06,ralph05,Donhauser01,Dujardin05,Wu04,ElbingOKFHWEWM05,cavin06,choi06,MorescoMRTGJ01,dulic03}.
Azobenzene, a molecule that changes conformation as a result of a
\textit{trans}/\textit{cis} transition when exposed to radiation,
could form the basis of a light-driven molecular
switch\cite{HugelHCMSG02,ZhangDCZRK04,Ratner06}. It is therefore
crucial to clarify the electrical transport characteristics of this
molecule. Here, we investigate theoretically charge transport in a
system in which a single azobenzene molecule is attached to two
carbon nanotubes. In clear contrast to gold electrodes, the
nanotubes can act as true nanoscale electrodes and we show that the
low-energy conduction properties of the junction may be dramatically
modified by changing the topology of the contacts between the
nanotubes and the molecules, and/or the chirality of the nanotubes
(that is, zigzag or armchair). We propose experiments to demonstrate
controlled electrical switching with nanotube electrodes.}

Recent theoretical studies\cite{ZhangDCZRK04,Ratner06} have explored
the use of azobenzene as an electronic switch when contacted by Au
electrodes. Promising results were obtained, as a significant change
in conductance was seen between the two conformations. Although a
mechanism was proposed that allowed for a certain movement of the
Au-leads, azobenzene and similar molecules undergoing isomerization
transitions require more versatile nanocontacts that adjust
appropriately to the change of length in going from one isomeric
state to the other. Moreover,
 the coupling to Au electrodes only leads
to a broadening of the molecular orbitals, without essentially
perturbing the intrinsic molecular electronic structure. Although
this may be a positive feature in some situations, more exciting is
the possibility of having nanoscale electrodes with electronic
properties  that may lead, in contrast to Au electrodes, to a strong
modification of the low-energy molecular electronic structure and
hence to dramatic changes in conductance. The most attractive
candidates are carbon nanotubes (CNTs)\cite{GuoSKWPTHCHOYBWHKN06},
which are essentially one-dimensional systems whose conduction
character (metallic or semiconducting) can be easily tuned by
changing their chirality.

We focus in this paper  on the transport characteristics of single
\textit{trans} and \textit{cis} azobenzene contacted by metallic
CNTs with two typical chiralities: armchair and zigzag. In order to
improve the molecule-electrode coupling, the hydrogen atoms at the
para-positions on the molecules are substituted by NHCO-groups,
which have been used as linkers to CNT electrodes in recent
transport experiments (A. Holleitner, private communication). Our
main aims are to highlight the dependence of charge transport on the
electronic structure of the nanotube electrodes, as well as the
influence of the linker groups, which are known to play an important
role\cite{steigerwald06}. It turns out that the oxygen atom plays a
critical role in determining the conductance near the Fermi level.
Finally, the sensitivity in the electrical response of the isomers
may be exploited to identify them using electrical transport
measurements as well as to realize a molecular switch.

%------------------------------------------------------------------------%
\begin{figure}[b]
\pippo{\includegraphics[width=.999\linewidth]{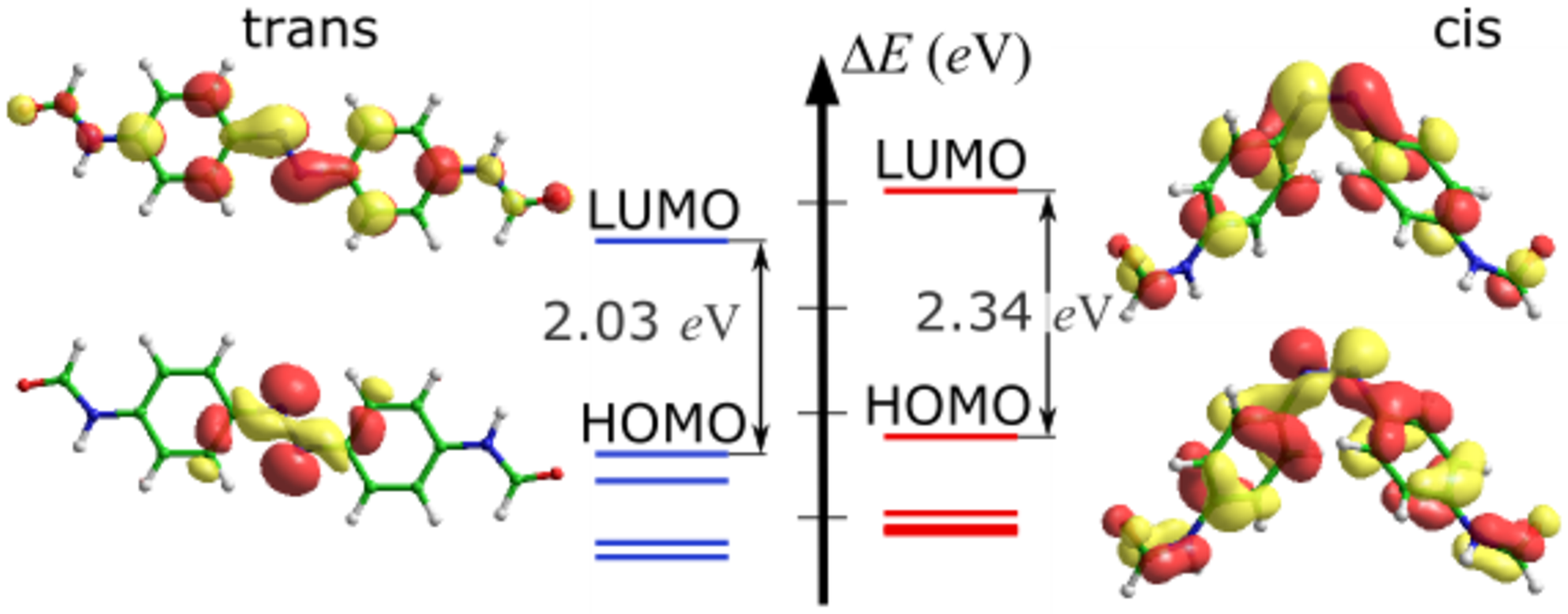}}
\myCaption{\label{fig:isolated} Low-energy electronic states of
azobenzene including the NHCO linkers.}{The positions of the
frontier molecular orbitals of the \textit{trans} (left) and
\textit{cis} (right) isomers are shown. The HOMO-LUMO gap is
slightly smaller in the \textit{trans} state, which is also the
lowest-energy configuration. The presence of the linkers does not
appreciably modify the geometric structure of the isomers. Also
shown is a density isosurface plot for the HOMO and LUMO orbitals.
The \textit{trans}-HOMO is mainly localized around the N-N dimer
with almost no weight on the linker groups. Notice that an extra
hydrogen saturates the linker. On attaching the molecules to the CNT
electrodes, this H-atom is removed and plays no further role.}
\end{figure}
%------------------------------------------------------------------------%

The \textit{trans} isomer is characterized by a planar structure,
but in the \textit{cis} isomer the benzene aromatic rings are tilted
with respect to each other, as seen in Fig.~\ref{fig:isolated}.
Total energy calculations yield an energy difference between the
isomers of $\sim 0.2\ e\textrm{V}$, the \textit{trans} state having
the lowest energy and smallest highest occupied molecular orbital
(HOMO)/lowest unoccupied molecular orbital (LUMO) gap of  $\sim
1.98\ e\textrm{V}$. On addition of the linkers, the gap slightly
increases to $ 2.03 \ e\textrm{V}$. Conversely, the gap for the
\textit{cis} isomer decreases from $ 2.55 \ e\textrm{V}$ to  $ 2.34
\ e\textrm{V}$. However, the isomer geometries, when the linkers are
added, are not appreciably modified after relaxation. In
Fig.~\ref{fig:isolated} we show the energetic position of the
isomers' frontier orbitals  as well as the optimized geometries.

%------------------------------------------------------------------------%
\begin{figure}[t]
\pippo{\includegraphics[width=.999\linewidth]{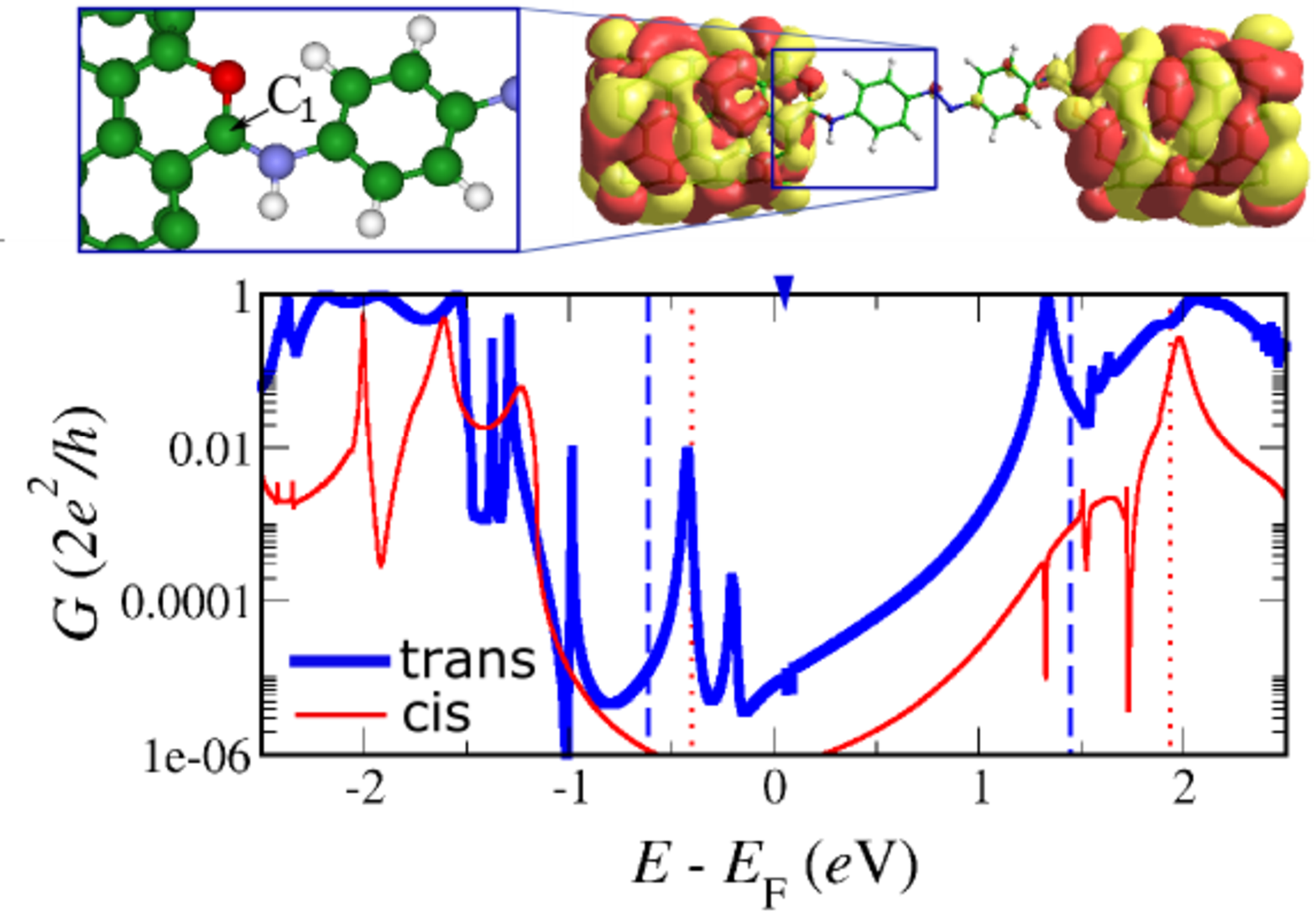}}
\myCaption{\label{fig:55CNT} Linear conductance at zero bias for a
(5,5) armchair-\textit{trans}(\textit{cis})-armchair junction.}{As a
reference we indicate by vertical lines the  positions of the HOMO
and LUMO orbitals of the isolated \textit{trans} (blue dashed lines)
and \textit{cis} (red dotted lines) molecules. All energies are
shifted to set the Fermi energy at $ 0 \ e\textrm{V}$, as only
energy differences from the Fermi level are revealed in transport
observables. Top panel: density isosurface plot of a selected
molecular orbital of the extended \textit{trans} molecule (at an
energy indicated in the conductance plot by the arrow near the Fermi
level). The low weight on the molecule reflects the rather low
conductance at this energy. The inset is an enlargement of the
CNT-molecule interface showing the formation of a carbon hexagon
with an oxygen defect on relaxation. This configuration is the same
for both isomers. Note that the carbon atom denoted by C$_{1}$ forms
a "bottleneck" in the junction, which means that the conductance is
largely determined by the spectral weight on the O-C$_1$ bond.}
\end{figure}
%------------------------------------------------------------------------%

Let us consider  the CNT-azobenzene-CNT molecular junction. The
complexity of the system leads to a very structured potential-energy
hypersurface with  many metastable states; for the sake of
simplicity we only consider coaxial arrangements of the left and
right CNTs. Different junction geometries varying in the way the
molecule is attached to the CNT surfaces were then optimized using
conjugated-gradient techniques, and the lowest-energy configurations
were used for the transport calculations. We note at this point that
the energy differences between the  investigated metastable states
are roughly of the order of a few $\textrm{m}e\textrm{V}$ per atom,
so that our transport calculations can be considered as providing
expected trends. Other configurations close in energy may lead to
slightly different quantitative results. However, the basic
conclusions of this paper should not be affected.

%------------------------------------------------------------------------%
\begin{figure}[t]
\pippo{\includegraphics[width=.999\linewidth]{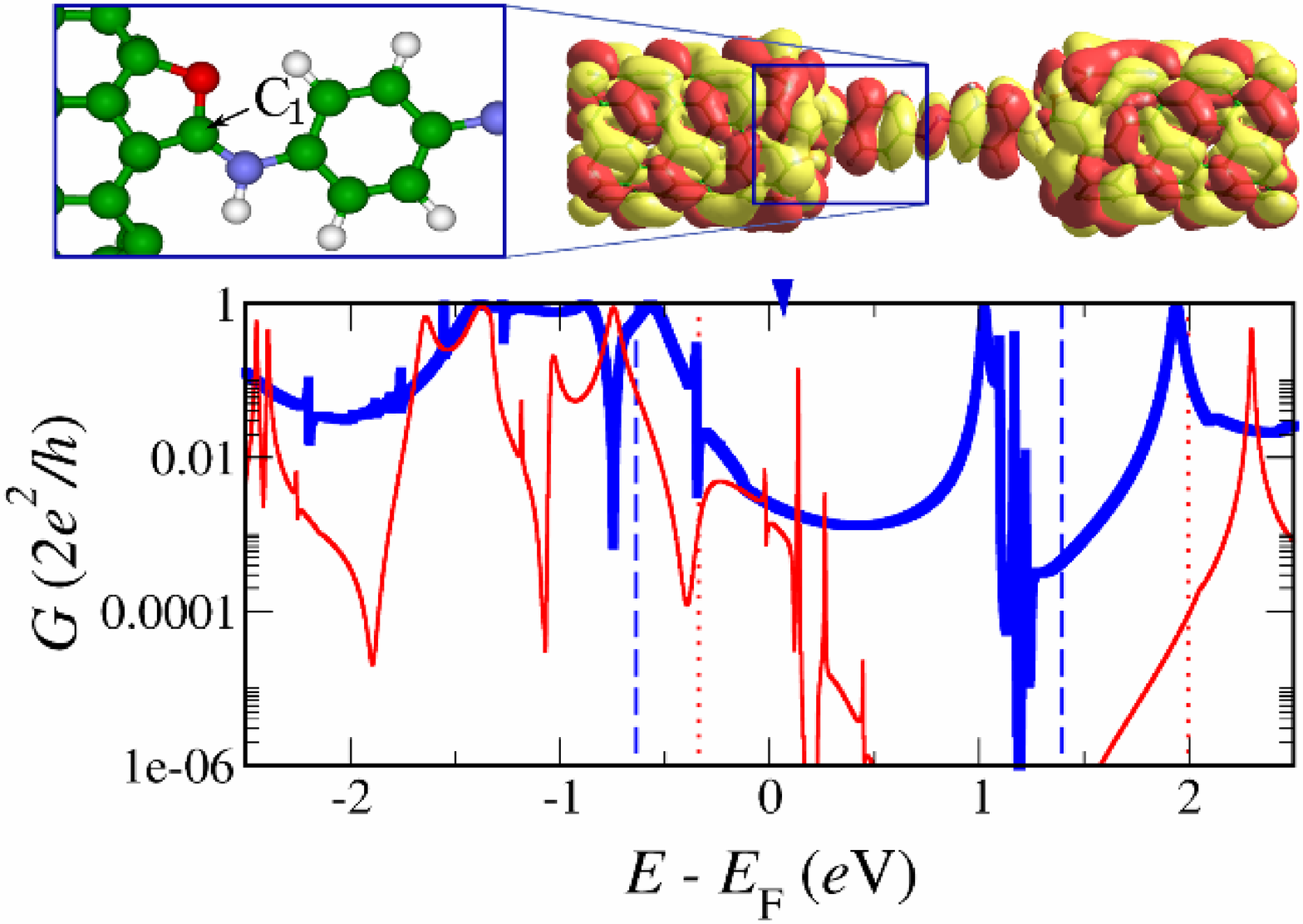}}
\myCaption{\label{fig:90CNT} Linear conductance at zero bias for a
(9,0) zigzag-\textit{trans}(\textit{cis})-zigzag junction.}{As in
Fig.~\ref{fig:55CNT}, the vertical dashed (dotted) lines indicate
the positions of the frontier orbitals of the isolated
\textit{trans}(\textit{cis}) isomers. Notice the increased
conductance around the Fermi energy for both isomers as well as the
more complex structure of the spectrum in this energy region. This
is closely related to the strongly localized edge state at the Fermi
level in zigzag tubes. Top panel: density isosurface plot of a
selected orbital of the extended \textit{trans} molecule (at an
energy indicated by an arrow in the conductance plot). Note the much
higher weight on the molecule when comparing with the armchair
junction in Fig.~\ref{fig:55CNT}, which explains the higher
transmission at this energy. The inset shows the local atomic
structure at the CNT-molecule interface, where a pentagon is formed.
This local bonding topology is similar for both isomers.}
\end{figure}
%------------------------------------------------------------------------%

Concerning the CNT-molecule contact, two direct covalent bonds are
formed between the azobenzene complex and the CNT leads: a C$_{\rm
CNT}$-C$_{1}$ bond and a C$_{\rm CNT}$-O bond. C$_{1}$ denotes the
carbon atom belonging to the linker. The bond topology is, however,
different for armchair and zigzag tubes, as shown in  the insets of
Figs.~\ref{fig:55CNT} and \ref{fig:90CNT}. In the former case a
hexagon is formed, and in the latter case a pentagon is built. This
appears to be a common feature of the bonding topology, as other
initial configurations led to similar connectivities upon
relaxation. In the \textit{trans} configuration attached to armchair
leads, one of the benzene rings slightly rotates out of plane by
about $8\g$ so that the relaxed state deviates from the planar
geometry found in the isolated molecule. This structural distortion
becomes even stronger when attaching the \textit{trans} molecule to
zigzag tubes ($14\g$). On the other hand, for the \textit{cis}
isomer, geometrical modifications on attachment to the leads do not
depend so much on the chirality of the CNTs, and consist essentially
of a slight reduction of the tilt angle.

Once the geometries were optimized, we first focused on the
dependence of the linear conductance on the charge injection energy
(Figs.~\ref{fig:55CNT} and \ref{fig:90CNT}) for armchair and zigzag
junctions, respectively. It turns out that the \textit{trans}
configuration shows an overall better ``transparency'' (higher
conductance)  than the \textit{cis} state, independently of the CNT
chirality, thus suggesting that the switching behaviour may already
be expected at low bias. The conductance of the
armchair-\textit{cis} junction displays a parabola-like gap around
the Fermi level, indicating that transport can only take place
through tunnelling, and thus very low currents are expected.
However, several resonances are seen for the \textit{trans} state
around the Fermi energy, which, despite their relative low
intensity, may give rise to resonant transport. These features arise
from the hybridization of surface states with molecular orbitals.
The resulting mixed states have a non-negligible overlap with the
CNT extended bulk states and thus can contribute to transport. The
fact that such states lie within a low-energy window around $E_{\rm
F}$ can make conductance manipulation more feasible because these
states will sensitively depend on the molecule-CNT contact topology,
as can be seen from our calculations. The low transmission of the
\textit{cis}-state is related to its distorted geometry, which
considerably breaks the $\pi$-conjugation along the molecular frame.

Turning now to the zigzag-azobenzene junctions, the conductance
appears to be considerably larger for both isomers than for armchair
junctions, as seen in Fig.~\ref{fig:90CNT}. One should notice at
this point that for both chiralities there are essentially two
interfering pathways  from the CNT to the azobenzene molecule, which
merge in the carbon atom belonging to the linker. Our calculations
show that the O-C$_{1}$ pathway completely dominates the charge
transport efficiency through the junction, the oxygen atom in the
NHCO linker hereby playing a crucial role. Because of its higher
electroaffinity, the oxygen tends to deplete the spectral weight on
the C$_{1}$ atom and thus acts as a chemical gate at the atomic
scale. This mechanism is very effective for armchair junctions, but
fails to have a dominant effect for zigzag junctions. Our
density-functional-theory-based calculations as well as a
$\pi$-orbital model hamiltonian show that, contrary to the armchair
system, in zigzag junctions a much larger spectral weight on the
C$_{1}$ atom persists for energies around the Fermi level. The
reason for this is that the surface density of states of a zigzag
tube has, as is well known, a large spectral weight close to the
Fermi energy (corresponding to a localized state at the zigzag
edge)\cite{Fukui05}. Moreover, the strong edge resonance manifests
itself in the electrode self-energies $\Sigma_{\rm L/R}(E)$ and
leads to a rather complex modification of the lower-lying molecular
 electronic states which is absent for armchair junctions.
Thus, the conductance difference for both junctions  is a result of
the interplay between local (interface) chemistry and electrode
surface electronic structure.

%------------------------------------------------------------------------%
\begin{figure}[t]
\pippo{\includegraphics[width=.999\linewidth]{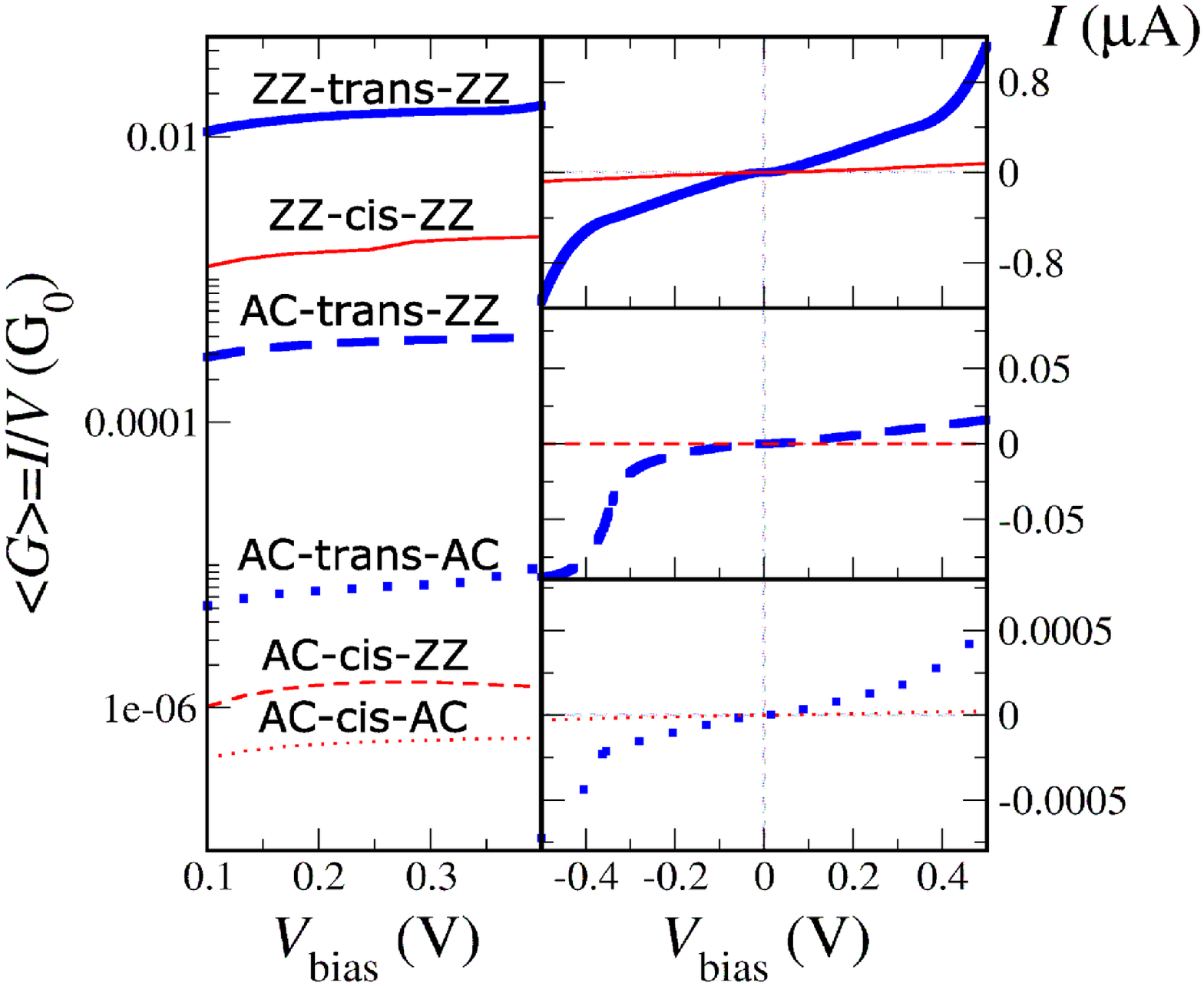}}
\myCaption{\label{fig:IV} Average conductance and current-voltage
characteristics of the CNT-azobenzene junctions.}{Three possible
CNT-molecule junctions are shown:
zigzag(ZZ)-\textit{trans/cis}-zigzag,
armchair(AC)-\textit{trans/cis}-zigzag, and
armchair-\textit{trans/cis}-armchair. A systematic hierarchy can be
seen, showing that the \textit{trans} state is always better
conducting than the \textit{cis} one and that armchair CNTs are
always detrimental to junction conductivity. }
\end{figure}
%------------------------------------------------------------------------%

To round off our discussion, we have calculated the current-voltage
($I$-$V$) characteristics of the junctions to confirm the possible
switching behaviour already suggested by the analysis of the linear
conductance. To reveal in a clearer way the effects of chirality on
transport, a hybrid armchair-azobenzene-zigzag system was also
investigated. The results for the three cases are displayed in
Fig.~\ref{fig:IV}. Charge transport for the \textit{cis} state is
strongly suppressed already at low bias for all junctions,
indicating a switching behaviour. Note that the
zigzag-\textit{trans}-zigzag junction shows up to three orders of
magnitude larger currents than the other two junctions, where
armchair tubes were used. This is even more dramatically seen in the
left panel of Fig.~\ref{fig:IV} where the average conductance $I/V$
is shown. These results further support the conclusion that the
electronic structure of both molecule and electrodes is crucial in
determining the low-voltage transport when nanoscale electrodes are
used. Another remarkable point is the strong rectifying behaviour
found in the hybrid armchair-\textit{trans}-zigzag junction. The
obvious chirality dependence of the effect suggests a different
mechanism to realize diode-like functions, which would not be
directly related to intrinsic electronic structure asymmetries of
the molecule as in the paradigmatic case of the Aviram-Ratner
molecular rectifier\cite{AviramR74}. We would further like to
mention that our equilibrium treatment of transport is validated by
recent non-equilibrium calculations\cite{Ratner06} using Au
electrodes which showed a similar qualitative behaviour for the two
isomers, that is,~$I_{\textit{trans}}(V)\gg I_{\textit{cis}}(V)$.

Finally, we would like to comment on possible experimental
realizations of the conformational switching when using CNT
electrodes. One precondition to achieve it is to increase the
flexibility of the junction, either by using longer linkers (which
should not, however, be detrimental to conduction by inducing, for
example, tunnel barriers) or by enlarging the electrode mobility.
One should keep in mind that the size of the spatial gap where the
molecule is placed is considerably smaller than the length of the
CNTs. Thus, small lateral variations in the position of the CNTs can
lead to a large change in the inter-electrode spacing. Another
possibility would lie in using telescopic CNTs as electrodes and
attaching the molecule to one of the inner shells\cite{zettl00}.
Additionally, we might expect that for a kinematically constrained
molecule such as the one studied here, the ``standard''
isomerization pathway would in principle be difficult to realize.
Although this might be true for bulk Au electrodes, the
one-dimensional nature of CNTs can considerably increase the
mechanical flexibility of the junction. Alternative pathways through
other metastable states compatible with the constrains can therefore
become effective. Indeed, several of the relaxed junctions displayed
geometric characteristics lying between those of the \textit{trans}
and \textit{cis} states. Light irradiation or mechanical actions
could trigger the transition from these metastable configurations to
the final \textit{cis} state. Another advantage of CNT electrodes
lies in the fact that they may posses a negative thermal expansion
factor\cite{Tomanek04}. A combination with other metallic electrodes
may thus lead to thermally more stable junctions, because normal
metal electrodes would expand upon heating and hence reduce the
junction flexibility.

\medskip%%

\fontsize{7.5}{9}\selectfont \noindent \textsf{\textbf{METHODS}}\\
We have used a density-functional-tight-binding (DFTB)
methodology\cite{FrauenheimSEHJPSS00} combined with Green function techniques\cite{PecchiaC04}
for geometry optimization and  transport calculations
 in the frame of the Landauer approach. The linear conductance $G$ can be obtained from the
quantum mechanical transmission probability at zero voltage:
$G=G_{0}T(E,V=0)$ (where $G_{0}=2e^2/h$ is the conductance quantum,
$e$ is the electron charge and $h$ is the Planck constant). The
transmission probability is related to the molecular Green function
$g(E)$ and the electrodes' self-energies $\Sigma_{\rm{L/R}}(E)$ by
$T(E,V)=4{\rm{Tr}}[g^{\dagger}(E) \ImP \Sigma_{\rm{R}}(E) g(E) \ImP
\Sigma_{\rm{L}}(E)]$ (ref.\onlinecite{FisherL81}). We stress that
when calculating the current $I(V)$, the full voltage-dependent
$T(E,V)$ is  used: $I(V)=(2e/h) \, \int dE \, \left[ f_{\rm
L}(E)-f_{\rm R}(E) \right] T(E,V) $. In the last expression
$f_{\rm{L/R}}(E)$ are Fermi functions of the left and right
electrodes.  $T(E,V)$ is calculated for the so called extended
molecule, which contains the azobenzene molecule, the linkers, and
the electrodes' surface atoms. This allows the molecule-electrode
coupling and the molecule electronic structure to be treated on the
same footing\cite{GrossmannGS02a,PecchiaC04}.

\fontsize{7}{10}\selectfont\sf%

%% Define the addendum environment for Supplementary Info, Acknowledgements, etc.
\newenvironment{addendum}{%
 \setlength{\parindent}{0in}%
\fontsize{7}{10}\selectfont\sf%
 \begin{list}{\textsf{Acknowledgements}}{%
 \setlength{\leftmargin}{0in}%
 \setlength{\listparindent}{0in}%
 \setlength{\labelsep}{0em}%
 \setlength{\labelwidth}{0in}%
 \setlength{\itemsep}{12pt}%
 \let\makelabel\addendumlabel}
 }
 {\end{list}\normalsize}

\newcommand*{\addendumlabel}[1]{\textbf{#1}\hspace{1em}}
\begin{addendum}
 \item
We thank A. Holleitner for making us aware of the details of the
experiments being performed with azobenzene, and N. Nemec, D.
Tomanek, and R. de Vivie-Riedle for discussions and suggestions.
This work was partially funded by the Volkswagen Foundation under
grant No.~I/78~340, by the DFG Priority Program ``Quantum Transport
at the Molecular Scale'' SPP1243, by the MEC under contracts
MAT2005-01388, NAN2004-09109-CO4-04, by the CAM under contract No.
S-0505/ESP-0200, and by the European Union project ``Carbon nanotube
devices at the quantum limit'' (CARDEQ) under contract No.
IST-021285-2. MdV acknowledges the support from the FPI Program of
the Comunidad Aut\'{o}noma de Madrid.
 \item[Competing Interests] The authors declare that they
have no competing financial interests.
 \item[Correspondence] Correspondence and requests for
materials should be addressed to GC~(email:
G.Cuniberti@physik.uni-R.de).

\clearpage
\item[Supplementary Information]
 To get further intuition on the enhanced
conductance in the zigzag-azobenzene junction, we have formulated a
minimal single-orbital model (see Fig. S1). The model consists of a
semi-infinite armchair (zigzag) CNT, a dimer which mimics the
C$_1$-O subsystem at the CNT-molecule interface, and a semi-infinite
chain connected to the C$_1$ site, which only fulfills the function
of acting as a charge reservoir and will subsequently be treated
within the wide-band approximation, \textit{i.e.} assuming an
energy-independent coupling between the dimer and the chain.

With this model, we can \textit{qualitatively} describe the local
chemistry and topology at the CNT-azobenzene interface. Our main
goal is to highlight the influence of the charge state at the
``oxygen'' site --which can be controlled by variations of its
onsite energy $\varepsilon_O$-- as well as the role played by the
nanotube surface electronic structure.

We can write an analytical expression for the transmission function
in terms of the dimer Green function: \beann T(E)= 4\Gamma_R \lcb
\Gamma_L^{11}(E) \lsb |G_{C_1C_1}(E)|^2 + |G_{C_1O} (E)|^2 \rsb +\rnothing\\
\lnothing 2\Gamma_L^{12}(E) Re(G_{C_1C_1}(E) G_{C_1O}^*(E))\rcb.
\eeann

\setcounter{figure}{0}
\renewcommand{\figurename}{Figure S}
%------------------------------------------------------------------------%
\begin{figure}
\pippo{\includegraphics[width=.9\linewidth]{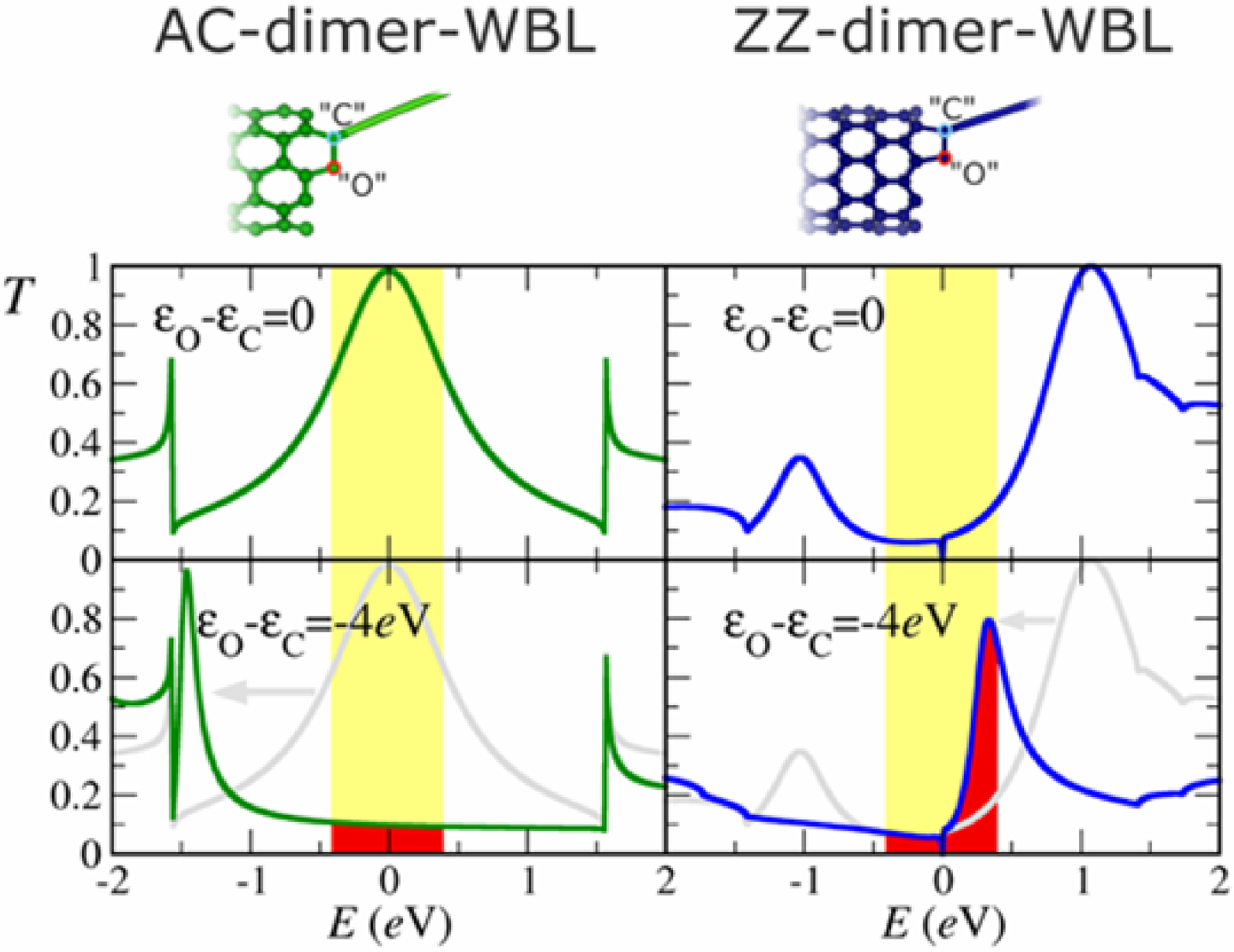}}
\myCaption{Transmission spectrum for the $\pi$-orbital model.}{The
left panels refer to a (5,5) armchair-C$_1$O dimer-wide band lead
(WBL) junction and the left ones to the case in which the nanotube
is a (9,0) metallic zigzag tube. The upper panels refer to a full
carbon junction with equal onsite energies on the dimer. In the
lower panels the onsite energy of the atom indicated as O in the
sketches of the junctions has been lowered to mimic the different
electronic affinities of carbon and oxygen. The shaded yellow area
refers to the integration region of $0.8$ Volt over which we have
calculated the $I$-$V$ characteristics in the main text. As it can
be clearly seen in the lower panel, the zigzag junction transmission
profile does result in a much higher current than for the armchair
one inside this voltage window.}
\end{figure}
%------------------------------------------------------------------------%
In this equation, $\Gamma_L^{11}(E)$ and $\Gamma_L^{12}(E)$ describe
the direct and cross coupling of the dimer to the left CNT electrode
(the indices 1 and 2 refer to the C$_1$ and O atoms, respectively),
and $\Gamma_R$ the constant coupling of the C$_1$ atom to the
wide-band lead (WBL) electrode on the right. The dimer Green
function can be obtained as solution of a Dyson $2x2$ matrix
equation $G^{-1}(E) = E-H_{dimer}- \Sigma_L(E)- \Sigma_R(E)$.

Three contributions to the transmission function can be clearly
distinguished: $G_{C_1C_1}$ is related to tunnelling pathways going
from the CNT directly through the C$_1$ atom, $G_{C_1O}$ describes
pathways passing through the O atom and then through the C$_1$ atom,
and the last term describes interference effects between both
pathways. Our calculations show that $|G_{C_1O}(E)|^2$ is the
dominant contribution to $T(E)$ for energies around the Fermi level.
Additionally this Green function matrix element is the most strongly
affected by a shift of the O onsite energies. Thus, we can
approximately write: $T(E\approx E_F) \approx 4 \Gamma_R
\Gamma_L^{11} |G_{C_1O}(E)|^2$.

In Fig. S1 we show the influence of a shift in $\varepsilon_O$ on
the transmission function for both armchair and zigzag junctions (we
take the C$_1$ onsite energy equal to that of the CNT carbon atoms).
It turns out that changes in $\varepsilon_O$ lead to a dramatic
suppression of the transmission for armchair junctions at low
energies (left panel of Fig. S1). For the zigzag junction, on the
contrary, electronic states lying slightly above the Fermi energy
are shifted down (chemical gating) towards $E_F$ without overtaking
it, and thus enter the energy window relevant for the low-bias
transport (right panel of Fig. S1). The value  $\varepsilon_O= -4
e\textrm{V}$ has been extracted from our DFTB calculation. It is the
difference between the $2p$ orbital energies of isolated oxygen and
carbon atoms. The effect discussed here is nevertheless stable
against variations around this value.

Extended calculations, which are not included here, show that this
peculiar behavior of the zigzag junction should be attributed to the
well-known surface state at the Fermi energy: with zigzag (ZZ) CNT
leads the dimer resonances are ultimately shifted (renormalized) by
the real part of the surface self-energy, which can, to a good
approximation, be written near the $E_F$ as:
$Re\Sigma^{ZZ}(E)\propto(E-E_F)/(\delta^2+(E-E_F)^2)$, inducing
\textit{de facto} a pinning of such resonances around the Fermi
energy, where the surface state exists with a linewidth $\delta$. In
this way the dimer eigenstates shift into a transport sensitive
energy region.

\end{addendum}

\end{document}